# AB INITIO STUDIES OF ELECTRONIC EXCITATIONS IN REAL SOLIDS

Adolfo G. Eguiluz and Wei Ku

Department of Physics and Astronomy, The University of Tennessee, Knoxville, TN 37996-1200

and Solid State Division, Oak Ridge National Laboratory, Oak Ridge, TN 37831-6030

## ABSTRACT

The first part of this article centers on the fact that key features of the dynamical response of weakly-correlated materials (the alkalis, Al), have been found experimentally to differ qualitatively from simple-model behavior. In the absence of *ab initio* theory, the surprises embodied in the experimental data were imputed to effects of dynamical correlations. We summarize results of *ab initio* investigations of linear response, performed within time-dependent density-functional theory (TDDFT), in which the unexpected features of the observed spectra are shown to be due to band-structure effects. For example, the key process behind the anomalous dispersion of the plasmon linewidth in K turns out to be decay into electron-hole pairs involving flat *d*-bands; the much-discussed double peak in the dynamical structure factor of Al is traced to a final-state gap. These results entail a new way of thinking about the physics of the electronic response in these systems. Contrary to conventional wisdom, the response cannot be understood "universally," in terms of a simple scaling with the density, on going from metal to metal (e.g., through the alkali series) — even the *shape* of the dispersion curve for the plasmon energy is system-specific. The second part of this article starts out with the observation that a similar *ab initio* study of systems with more complex electronic structures would require the availability of a realistic approximation for the dynamical many-body kernel $f_{xc}$ entering the density-response function in TDDFT. In its absence, we outline a diagrammatic alternative, framed within the conserving-approximation method of Baym and Kadanoff. Using as a benchmark the band gap of Si obtained in the *GW* approximation, together with a diagrammatic (and conserving) solution of the ensuing Bethe-Salpeter equation (which includes dynamically-screened particle-hole ladders and other fluctuation processes), we discuss issues involving conservation laws, self-consistency, and sum rules. These conceptual issues are particularly important for the development of *ab initio* methods for the study of dynamical response and quasiparticle band structure of strongly-correlated materials. We argue that inclusion of short-range correlations absent in the *GW* approximation is a must, even in Si.

# INTRODUCTION

The current theoretical understanding of the electronic correlations, and their impact on the properties of materials, is based, for the most part, on simple-model descriptions. In broad terms, the available models derive from two extreme pictures of a solid. On the one hand we have the homogeneous electron gas (or jellium) model, usually invoked in the study of simple metals.[1] In this model, the dominant excitation for small wave vectors is the plasmon, whose dispersion relation has traditionally been assumed to be well described by the long-range interactions contained in the random-phase approximation (RPA).[1] Because of the success of the local-density approximation (LDA) for the exchange-correlation energy functional introduced in Kohn-Sham density-functional theory,[2] the jellium model turns out to play an important "supporting role" in modern electronic structure calculations. On the other hand, we have the Hubbard model,[3,4] originally introduced for the study of itinerant magnetism in the $3d$ metals; this model emphasizes short-range aspects of the correlations, as it allows the electrons to interact only when they encounter at a given lattice site. In both extreme pictures the description of the band structure is reduced to a minimum: In jellium, it is ignored altogether (more precisely, there is one parabolic band); in the Hubbard model, it is represented by one (sometimes three, in the context of the copper-oxide superconductors) tight-binding bands.

The central theme of this article is the physics of the electronic excitations in real materials. This is a vast subject. The first part of our presentation is devoted explicitly to systems which, on account of their relative simplicity, have recently become amenable to *ab initio* theory. We start out with the observation that, even for jellium-like, weakly-correlated materials, such as the alkali metals and Al, high-resolution spectroscopic measurements[5-9] have uncovered significant "anomalies," whose explanation requires a new way of thinking about the physics of the excitations. For example, the dispersion of the plasmon *energy* in Cs turns out to be negative for small wave vectors, contrary to the case of, say, K, in which the dispersion is positive, as expected from textbook physics,[1] and general considerations of stability. Furthermore, the dispersion of the plasmon *linewidth* in K is positive,[5] contrary to the results of a simple, intuitive, model of the dielectric response which relies on the assumed weakness of the crystal potential (a negative plasmon linewidth dispersion is a "generic" feature for the alkalis series in this model).[10] In addition, the overall shape of the dynamical structure factor of Al (and other systems such as Be, Li, Si, and graphite)[7-9] is dominated, for large wave vectors, by a two-peak structure whose existence is beyond RPA theory for the homogeneous electron gas.

The above phenomenology is quite remarkable. The fact that experiment tells us that from the knowledge of the dispersion curve for the plasmon energy in, say, K, one cannot predict the *shape* of the dispersion curve in Cs, clearly calls for new theoretical insight into the physics of correlated electrons in these systems. A similar message follows from the qualitative deviations from simple-model predictions observed in the loss spectrum of Al and other systems. Such insight may be sought within two scenarios:

*(i)* We must abandon the long-cherished idea that the excitations in these materials can be understood "universally," i.e., in terms of a simple scaling, controlled solely by the density parameter $r_S$, on going from system to system. This viewpoint suggests that one must go beyond simple-model descriptions of the underlying band structure.

*(ii)* Scaling with the density may be restored if we assume that what breaks down is scaling à la RPA; in this case, then, for example, there must be some threshold value of $r_S$ in the range 4.8 –5.6 (corresponding to K and Cs, respectively) for which strong non-RPA correlations must emerge, and be responsible for switching the plasmon energy dispersion from positive to negative. This viewpoint is commonly pursued together with the premise that the effects of the band structure are not important.

In our view, to date there is no model of the correlations that is able to reproduce the experimental facts convincingly within scenario *(ii)*. In this article, whose scope is that of a pedagogical overview of some of our ideas about a rapidly-developing field, we show that scenario *(i)* leads to a compelling interpretation of the available experimental information on the basis of *ab initio* calculations of the excitation spectra.[11-20]

We consider first the plasmon in the alkali metals. Interestingly, in the absence of *ab initio* theory, it was suggested that the "anomalies" in the excitation spectra of K and Cs, observed via electron energy-loss spectroscopy (EELS) measurements,[5,6] must be due to unspecified dynamical short-range correlations —which falls in the realm of *(ii)*. We compare calculated and measured dispersion curves for the plasmon energy in Na, K, and Cs. Our results illustrate vividly the absence of conventional RPA scaling with the density. Band-structure effects account for the clear-cut breakdown of simple-model physics contained in the data; these effects are mainly due to the presence, in both K and Cs, of *d*-bands for energies comparable with that of the collective mode. Next, we show that the key mechanism behind the anomalous dispersion of the plasmon linewidth in K is decay into particle-hole pairs involving empty states of *d*-symmetry.[20] In all these calculations, quantitative agreement with the data is achieved within scenario *(i)*.

We also address the dynamical structure factor of Al, a notorious example of the duality of outlook implied by the above two scenarios. Under the simplifying assumption that "Al is jellium," most of the many papers published on this problem over the years sought to explain the two-peak loss structure originally observed via inelastic x-ray scattering spectroscopy (IXSS) measurements[21,22] in terms of dynamical correlations. For recent attempts within this approach, which is consistent with *(ii)*, the reader is referred to papers by Platzman et al.[7] and Schülke et al.[8] Here we sketch results of *ab initio* calculations which lead to a "natural" explanation —along the lines of *(i)*— of the key features of the spectrum for large wave vectors.[12,15] The two-peak structure is shown to correlate strikingly with a wide quasi-gap in the density of states (DOS) of Kohn-Sham single-particle states.

The density-response calculations underlying the results just referred to are based on a linear-response implementation of time-dependent density functional theory (TDDFT).[23,24]

In the TDDFT linear-response scheme the many-body effects enter the loss spectrum in two ways: via a dynamical many-body kernel, $f_{xc}$ (loosely speaking, a vertex function), and through the effects of exchange and correlation included in the single-particle Kohn-Sham density-response function $c_s$. We note at the outset that the basic reason why we are able to elucidate the above-mentioned anomalies is that, upon switching off explicitly the dynamical correlations contained in $f_{xc}$, the calculated observables either agree nearly quantitatively with the data (e.g., the dispersion curve of the plasmon in the alkalis for small wave vectors), or reproduce the main qualitative features of the spectrum quite well (e.g., the loss function of Al for large wave vectors).

In this context, an interesting finding is the role played by the exchange-correlation effects built into the *unoccupied* Kohn-Sham states (which we treat in the LDA) in the explanation of the positive dispersion of the inverse lifetime of the plasmon in K. This novel result is traced to the fact that the damping process is dominated by a subtle interplay between flat *d*-bands and the collective mode.

Thus, for the relatively-simple systems alluded to above, the single-particle response $c_s$ provides a "starting point" which by itself contains a good deal of the physics of the electron-hole response as it manifests itself experimentally. However, for systems with more complex electronic structures (for example, transition metals, and their oxides) the dynamical correlations contained in $f_{xc}$ should play a significant role. Unfortunately, explicit representations for this kernel which would make possible the *ab initio* study of dynamical response in those systems are not yet available.[24] Thus, in the second part of this paper, we pursue a diagrammatic alternative within the Baym-Kadanoff method of conserving approximations.[25,26] In this approach, the counterpart of the TDDFT integral equation for the density-response function is the Bethe-Salpeter equation for the two-particle correlation function, whose kernel, the effective electron-hole interaction, depends on four coordinates, instead of just two, as $f_{xc}$ does. A diagrammatic solution of this equation within the shielded-interaction approximation (SIA) for the electron self-energy,[25] yields a formal result for the irreducible polarizability $P_{SIA}$ which includes, to all orders, the effects of shielded particle-hole ladders and other fluctuation diagrams which arise naturally in the conserving, functional-differentiation approach. The formal relationship between $f_{xc}$ and the (inverse) polarizability is touched upon, as it may suggest ways of building into the former the dynamical processes contained in the latter.

Now the road to the two-particle problem goes through the one-particle problem, i.e., the Dyson equation, which we address first. In the important case of silicon, the SIA yields a value of the absolute band gap which is too large, compared to experiment.[27] This conclusion brings the "band-gap problem" back to the fore, as the SIA is nothing but Hedin's *GW* approximation,[28] which was supposed to have solved the problem —but only in the absence of self-consistency. (Recent, extensive reviews of work performed within the *GW* approximation are available.[29,30]) Thus, a more elaborate self-energy must be considered. Within the conserving scheme, this may be done by augmenting the Baym $\Phi_{SIA}$ functional via the inclusion of short-range correlations —e.g., particle-hole and particle-particle ladders. In view of recent efforts to go beyond the *GW* approximation, and of related discussions of conservation laws and sum rules,[30-38] we address these issues in some detail, because of their significance for the development of meaningful approximations for the response function and quasiparticle band structure in correlated-electron systems.

**DYNAMICAL DENSITY RESPONSE IN TDDFT**

A general formulation of the time evolution of an interacting electron system in an external potential $v_e(\vec{x},t)$ has been given by Runge and Gross.[23] These authors established the invertibility of the mapping $v_e(\vec{x},t) \to n(\vec{x},t)$, where $n(\vec{x},t)$ is the time-dependent density for the interacting system. This result constitutes a time-dependent generalization of the Hohenberg-Kohn theorem[39] of ground-state density-functional theory.† We then have that the time-evolved many-particle state is uniquely determined (for the given initial state) as a functional of $n(\vec{x},t)$ —up to a purely time-dependent phase. Such phase is eliminated when evaluating mean values of physical operators for the many-body system.

A particularly important example is the quantum mechanical action, $A[n]$, which has a stationary point at the exact solution of the many-particle Schrödinger equation. From the previous argument it then follows that

$$\frac{\delta A[n]}{\delta n(\vec{x},t)} = 0 \;, \qquad (1)$$

i.e., the action is stationary at the exact density. Reasoning in a similar manner as done by Kohn and Sham[2] for the case of the ground state, from this variational principle one can prove[23] that $n(\vec{x},t)$ may be obtained within an equivalent single-particle picture, i.e., we can write

$$n(\vec{x},t) = \sum_n^{occupied} |\varphi_n(\vec{x},t)|^2 \;, \qquad (2)$$

in terms of the solutions $\varphi_n(\vec{x},t)$ of the time-dependent Kohn-Sham equation

$$i\hbar \frac{\partial}{\partial t}\varphi_n(\vec{x},t) = \left[ -\frac{\hbar^2}{2m}\nabla^2 + v_s[n](\vec{x},t) \right]\varphi_n(\vec{x},t) \;, \qquad (3)$$

where the single-particle potential $v_s[n](\vec{x},t)$ contains, in addition to the external and Hartree potentials, the exchange-correlation potential $V_{xc}[n](\vec{x},t)$, whose *functional* dependence on the density is implied by our notation. (Note that this scheme assumes non-interacting *v*-representability for the time-dependent density.[40])

---

† There is a conceptual difference though, having to do with the dependence of the time-evolved state on the initial many-body state.[40]

The TDDFT formalism provides a convenient framework for the study of the *linear response* of a many-electron system to an external potential $\delta_e(\vec{x},t)$.[24] In general, we define the density-response function $\chi$ by the equation $\delta n = \chi \delta_e$. In TDDFT the linear change in density can also be calculated as $\delta n = \chi^{(s)} \delta_s$, where $\chi^{(s)}$ is the *single-particle* response for the *unperturbed* Kohn-Sham system. From Eqs. (2) and (3) $\delta_s$ can be related to $\delta_e$, and this leads us to the following integral equation for the response function:

$$\chi = \chi^{(s)} + \chi^{(s)}[v + f_{xc}]\chi , \qquad (4)$$

where $v$ is the bare Coulomb interaction, and the exchange-correlation kernel $f_{xc}$ is defined by the equation

$$f_{xc}[n](\vec{x}t;\vec{x}'t') = \frac{\delta V_{xc}[n](\vec{x},t)}{\delta n(\vec{x}',t')} , \qquad (5)$$

where the functional derivative is to be evaluated at the unperturbed density. We emphasize that Eqs. (4) and (5) are formally exact; all explicit effects of dynamical correlations are contained in $f_{xc}$.

The spectral representation for the single-particle response $\chi^{(s)}$ in terms of Kohn-Sham eigenfunctions and eigenvalues is of the usual form. For the case of a periodic crystal its Fourier transform with respect to the spatial coordinates takes on the form

$$\chi^{(s)}_{\vec{G},\vec{G}'}(\vec{k};\omega) = \frac{1}{V_N} \sum_{\vec{k}'}^{BZ} \sum_{j,j'} \frac{f_{\vec{k}',j} - f_{\vec{k}'+\vec{k},j'}}{E_{\vec{k}',j} - E_{\vec{k}'+\vec{k},j'} + \hbar\omega + i\hbar 0} \langle \vec{k}',j | e^{-i(\vec{k}+\vec{G})\cdot\hat{\vec{x}}} | \vec{k}'+\vec{k},j' \rangle$$

$$\times \ \langle \vec{k}'+\vec{k},j' | e^{i(\vec{k}+\vec{G}')\cdot\hat{\vec{x}}} | \vec{k}',j \rangle , \qquad (6)$$

where $V_N$ is the normalization volume, $\vec{G}$ is a vector of the reciprocal lattice, $j$ is a band index, and all wave vectors are in the first Brillouin zone (BZ). We evaluate Eq. (6) using Kohn-Sham states and eigenvalues obtained within the LDA for the functional $V_{xc}[n]$. In the Fourier representation of Eq. (6), Eq. (4) is turned into a matrix equation that we solve numerically.

The above outline of linear response within TDDFT was presented in order to emphasize the point that the Kohn-Sham response $\chi^{(s)}$ enters Eq. (4) *rigorously* —even if the individual states in the sum in Eq. (6) do not correspond to physical excitations (as are observed, e.g., in photoemission and inverse photoemission experiments). This issue is important for a proper interpretation of the results we discuss in the next two sections, where one of our aims is to sort out the effects of the one-particle band structure from those of dynamical many-body correlations; the latter are turned off by setting $f_{xc} = 0$.[†]

---

[†] For systems with more complex electronic structures we expect that, for $f_{xc} = 0$, theory would be sufficiently far off experiment that the success story of the next two sections would not be reproduced.

# PLASMONS IN THE ALKALI METALS; ENERGY AND LINEWIDTH DISPERSION

In Fig. 1 we compare our calculated dispersion curves for the plasmon energy in Na, K, and Cs, with the EELS data of vom Felde, Sprösser-Prou, and Fink.[5] The plasmon energy for a given wave vector $\vec{k}$ was read off the position of the main peak in the loss function $\text{Im } c_{\vec{G}=0,\vec{G}=0}(\vec{k};w)$ (in the alkalis the plasmon dispersion curve is entirely confined to the first BZ). We have set $f_{xc} = 0$ in Eq. (4). As discussed elsewhere,[20] the sampling of the BZ required in Eq. (6) is efficiently performed by carrying out the calculation on the imaginary frequency axis; the analytic continuation to the real axis is performed via Padé approximants. Since higher-lying core states turn out to affect the plasmon dispersion and lifetime of the alkali metals heavier than Na, we evaluate $c^{(s)}$ using the full-potential linearized augmented plane wave (LAPW) method.[41] Also shown in Fig. 1 are the dispersion curves for electrons in jellium with the density corresponding to each metal (solid lines); note that, in jellium, the response $c$ obtained from Eq. (4) for $f_{xc} = 0$ is identical to the RPA response.

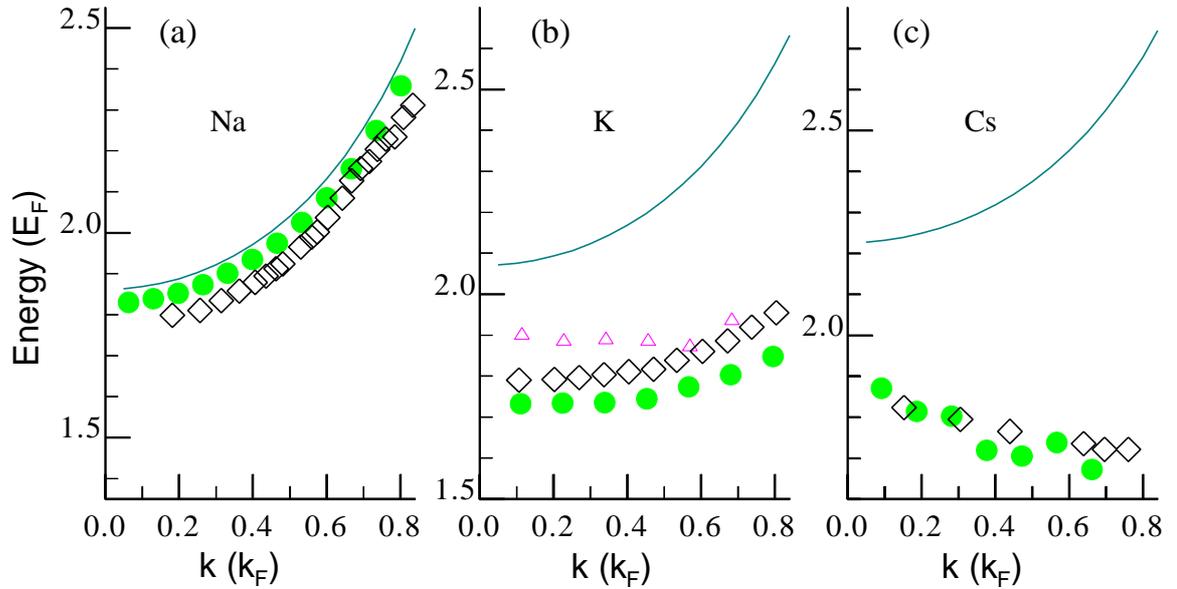

**Figure 1.** Absence of "universal" behavior, or RPA-like scaling, in the excitations in the alkali series. Shown are our calculated (solid circles) and measured[5] (empty diamonds) dispersion curves for the plasmon energy in Na, K, and Cs. Note: we have set $f_{xc} = 0$ in Eq. (4). The solid lines are the corresponding results for electrons in jellium. The empty triangles in panel (b) correspond to the dispersion curve for the K plasmon obtained upon excluding the contribution to Eq. (6) from the core states. (Theory is for (1,1,0) propagation; the EELS data[5] are for polycrystalline samples.)

From a glance at Fig. 1 it is apparent that there is very good agreement between calculated (solid circles) and measured (empty diamonds) dispersion curves. This is a non-trivial test of *ab initio* theory, as this agreement includes the dramatic change in the shape of the dispersion curve which is realized on going from Na to Cs.[5,14] We emphasize that this change in the physics of screening taking place through the alkali series implies that the loss function $\text{Im } c_{\vec{G}=0,\vec{G}=0}(\vec{k};w)$ (which, for small $k$'s is dominated by the plasmon) does not scale with, nor is controlled solely by, the density parameter $r_s$. We remind the reader that this conclusion is consistent with scenario *(i)* put forth in the Introduction.

The "anomalous" dispersion of the plasmon energy in Cs, and the consequent lack of universal behavior evident in Fig. 1, are traced to significant departures of the LDA band structure from nearly-free-electron behavior for energies of the order of the plasmon energy (whose value for zero wave vector we denote by $w_p(0)$). In Na the band structure is essentially free-electron like in that energy range; consequently, the plasmon dispersion curve is similar to the result for jellium, as seen in Fig. 1a. However, in K, as we illustrate below when we discuss plasmon damping, $w_p(0)$ falls in a spectral region where the LDA band structure is dominated by *d*-bands.¶ As a result, the plasmon dispersion curve (Fig. 1b) differs appreciably from the "canonical" jellium result. The extreme case is Cs (Fig. 1c), for which a qualitative breakdown of simple-model physics is realized: the plasmon dispersion is negative, for small wave vectors.†

Let us expand a little on the interplay between band-structure effects and the calculated plasmon dispersion curves for K and Cs.‡ Consider first the case of K, shown in Fig. 1b, where the empty triangles correspond to a calculation in which we have excluded the contribution to $c^{(s)}$ from the core states. From that result, and by comparison with the dispersion curve for jellium, we can visualize that the large "real-metal" effects at play recognize two sources: the "polarization effect" of the core states (mostly the $3p^6$-derived states), *and the polarization effect of d-bands lying just above* $w_p(0)$¶ —the latter effect being responsible for the major part of the difference between the solid line (jellium) and the empty triangles (crystal calculation without the effect of the core).$

As for the case of the plasmon in Cs, the dispersion curve given by the triangles in Fig. 2a corresponds to a calculation in which we have kept only the lowest three valence bands in the evaluation of $c^{(s)}$—one occupied, two empty bands. (The core states are also kept.) We stress the fact that the slope of the dispersion curve so obtained is *negative*; indeed, it differs little from the slope of the "all-band" dispersion curve (solid circles) —the shift between both dispersion curves is weakly *q*-dependent. This test establishes that *the first* $ Note that the core states affect not just the plasmon energy —which shifts upwards, when the core is excluded, as seen in Fig. 1b— but also the slope of the dispersion curve, which actually becomes slightly negative in the absence of these states.

---

¶ Cf. Fig. 4a shown below.

† Our results of Fig. 1 agree qualitatively with those of Aryasetiawan and Karlsson;[14] however, there appear to be differences in the analysis of the results. For example, in our calculations for K the *d*-bands are already degenerate with the plasmon; similarly, we find that the *d*-bands responsible for the negative dispersion of the Cs plasmon lie appreciably *below* the plasmon energy.

‡ For brevity, and since such discussion would take us away from the main theme of this paper, in Fig. 1 we have ignored the crystal local fields, i.e., the non-diagonal elements of the matrix defined in Eq. (6). Now, in Na these effects are very small. In K they are small but non-negligible; their inclusion shifts the calculated dispersion curve upward, relative to the solid circles in Fig. 1b, leading to an even better agreement with the data. In Cs these effects are larger, a reflection of the importance of the $5p^6$ semicore states, which are quite shallow —they lie at ~ 11 eV below the Fermi energy. As we have found out in earlier work,[18] in this case the many-body kernel $f_{xc}$ cannot be ignored, even for $k \to 0$.

*two empty bands are responsible for the anomalous dispersion of the plasmon energy in Cs* —that is, for the key physical difference with a jellium-like description of the Cs plasmon. It is noteworthy that these bands are mostly of *d*-character (the partially-full lowest band is, of course, predominantly *s*-like); they are the origin of the sharp structure lying at ~ 1-2 eV above the Fermi surface in the DOS shown in Fig. 2b. Higher-lying valence bands, particularly the *d*-complex lying just above $w_p(0)$ (~ 3eV), account for the large polarization shift observed in Fig. 2a between the three-band dispersion curve and the all-band result.†

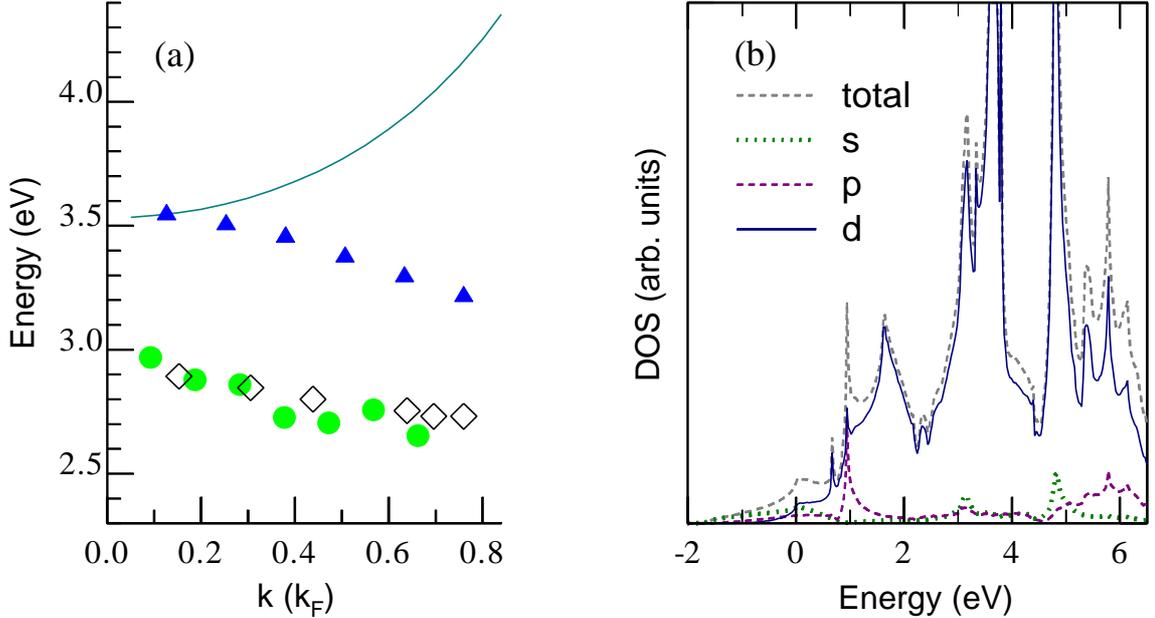

**Figure 2.** Physical explanation of the negative dispersion of the plasmon energy in Cs. (a): Plasmon dispersion curve (solid triangles) obtained upon keeping just the lowest three valence bands (plus the core states) in the evaluation of $c^{(s)}$. The solid circles, empty diamonds, and the solid line, are as in Fig. 1c. (b): LAPW-based DOS for Cs;[41] shown are the total DOS and contributions from states of *s*, *p*, and *d*, symmetry. The zero of energy is the Fermi energy. The spiky structure below 2eV corresponds to the first two empty bands kept in the calculation of (a). See text for details.

Finally, we note that the core states play an even more important role in the dynamical response of Cs than they do in the case of K. Indeed, in addition to being the source of a downward shift of the plasmon dispersion curve, the core states of Cs (mainly the $5p^6$-derived states, whose excitation threshold is at ~ 11 eV) have a profound impact on the *lineshape* of the plasmon loss. A related finding, which we will discuss elsewhere, is that the core states "screen out" the fine structure in the loss function due to interband transitions to states corresponding to the sharp peaks lying above $w_p(0)$ in the DOS shown in Fig. 2b.

---

† In previous work[18] we had concluded that the higher *d*-complex lying between 3eV and 4eV was responsible for the negative dispersion of the Cs plasmon. Figure 2a shows that, in agreement with the earlier analysis, this *d*-complex does bring about a large downward shift of the plasmon energy relative to the jellium result (this shift basically takes the triangles into the circles). However, the figure also shows that the effect of this complex on the *slope* of the dispersion curve is small.

Turning now to the physics of plasmon *damping*, Fig. 3a shows a well-converged calculation[20] of the dispersion of the plasmon *linewidth* in K (solid circles). The calculated full-width at half-maximum of the plasmon peak, $\Delta E_{1/2}(\vec{k})$, is compared with the value extracted by vom Felde et al.[5] from their EELS loss function (empty diamonds). (Note that Fig. 3 gives $\Delta E_{1/2}(\vec{k})$ relative to its value at zero wave vector.) Clearly, our result is in excellent agreement with experiment; since this agreement is obtained for $f_{xc} = 0$, we conclude that the plasmon linewidth *dispersion* of K is *not* controlled by a dynamical many-body mechanism.‡

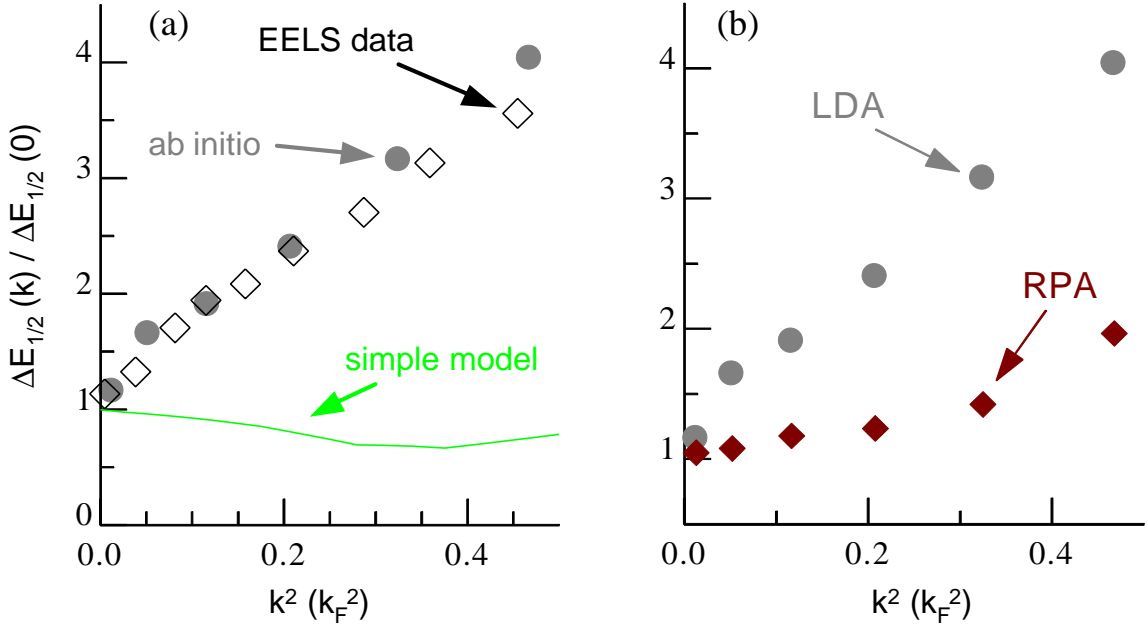

**Figure 3.** Physical explanation of the positive dispersion of the plasmon linewidth in K. (a): Comparison of calculated[20] (solid circles; labeled "*ab initio*") and measured[5] (empty diamonds; labeled "EELS data") linewidths. Note: we have set $f_{xc} = 0$ in Eq. (4). The solid line (labeled "simple model") corresponds to the calculations of Sturm and Oliveira.[10] (b): Plasmon linewidth dispersion obtained from Eq. (4) on the basis of RPA (solid diamonds) and LDA (solid circles) calculations of the single-particle response $c^{(s)}$. See text for details; cf. Fig. 4.

Our result is striking, as intuitive expectations based on the fact that the gap just above the Fermi surface at the *N*-point is small (cf. Fig. 4a) yield a $\Delta E_{1/2}(\vec{k})$ of the form shown by the solid line in Fig. 3a, which corresponds to an evaluation of the dielectric function to second order in the electron-lattice interaction[10] (modeled by a local, empirical pseudopotential). It is apparent that, although "physically reasonable," a simple-model description of plasmon damping in K breaks down, and badly so.

This breakdown can be understood by analyzing the impact of key "final state" bands, shown on Fig. 4a, in which the shaded strip is the energy interval representing all the single-particle states which may couple to the plasmon, as determined by the conservation laws of energy and crystal momentum. This analysis is done in detail elsewhere;[20] here we simply summarize results of several tests we performed in order to open up the "black box." For example, if we keep just the first three valence bands (thin solid lines in Fig. 4a) in the evaluation of $c^{(s)}$ according to Eq. (6), the calculated $\Delta E_{1/2}(\vec{k})$ happens to be quite comparable with the simple-model result displayed in Fig. 3a; this is reasonable, as the dispersion of these bands is quite substantial, even if they are not strictly parabolic.¶ Now, when three additional bands (thick solid lines in Fig. 4a) are included in $c^{(s)}$, the plasmon linewidth dispersion curve changes *qualitatively*; the corresponding $\Delta E_{1/2}(\vec{k})$ turns out to agree well with the "all band" calculation of Fig. 3a (and thus, with the EELS data as well). Clearly, then, these three bands provide the key decay channels for the K plasmon. Note that the point here is not the addition of three extra bands in the evaluation of $c^{(s)}$. What is crucial is that these bands are flat, and overwhelmingly of *d*-character.[20] The phase space complexity which these bands introduce in the plasmon damping process cannot be approximated via nearly-free-electron states —thus the anomalous behavior of the linewidth dispersion.

---

‡ This conclusion does not imply that the dynamical correlations vanish identically, but, rather, that the effects of $f_{xc}$ have little impact on the *dispersion* of the linewidth.

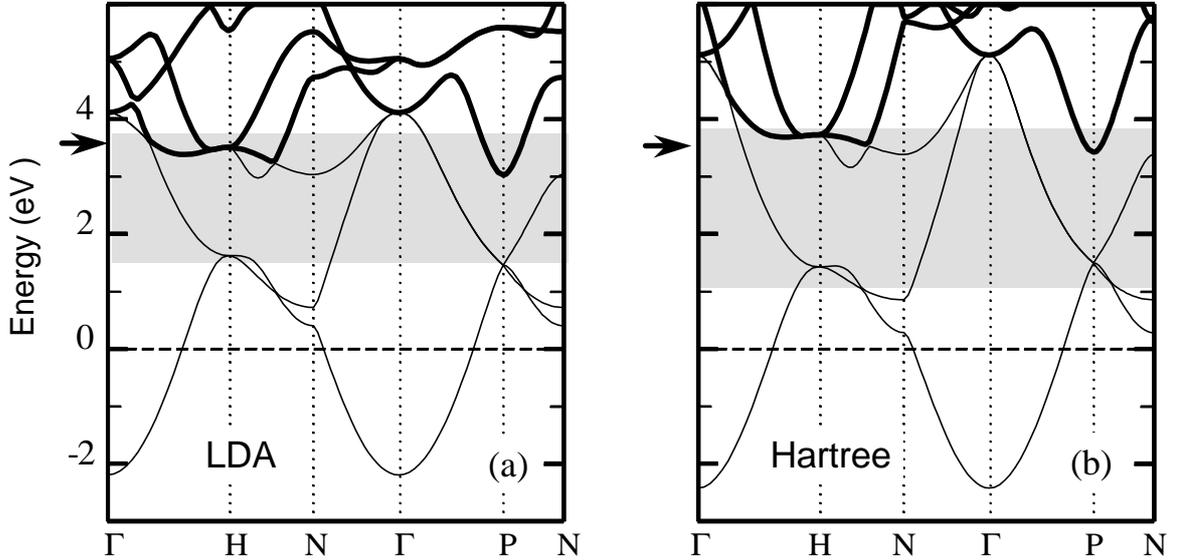

**Figure 4.** LAPW valence band structure of K for energies in the vicinity of the plasmon energy $w_p(0)$, whose value is indicated by an arrow. (a): LDA; (b) Hartree. The shaded strip illustrates, in each case, the energy location of all the possible empty states corresponding to electron-hole decay channels for the plasmon. See text for details; cf. Fig. 3.

We reiterate that the use of the Kohn-Sham states in Eq. (6) is "legal," and not a matter of expediency; the only approximation made in the evaluation of $c^{(s)}$ is the LDA. We pursue this issue further with reference to Fig. 3b, which shows results of additional calculations performed within the RPA, in which all exchange-correlation effects are left out, including those in the band structure, which corresponds to the Hartree approximation (i.e., we switch off $V_{xc}$ in the Kohn-Sham equation). The main physical change is that, as seen in Fig. 4b, the key flat bands are shifted upwards in relation to the LDA $d$-bands of Fig. 4a. Although the magnitude of the shift is not large, it does have an important impact on the linewidth, as the Hartree $d$-bands lie almost entirely above the corresponding shaded strip —thus, the decay channels involving these bands become inaccessible to the plasmon. This effect is the root of the reason why, as seen in Fig. 3b, the $\Delta E_{1/2}(\vec{k})$ calculated in RPA has a much smaller slope than the TDDFT result (and the EELS data).

---

¶ The lower edge of the shaded strip in Fig. 4 is given by (bottom of occupied band+$\hbar w_p(0)$); the upper edge is ($E_F + \hbar w_p(k_c)$), where $k_c$ is the wave vector for which the plasmon enters the continuum of particle-hole pair excitations; $\hbar w_p(0)$ =3.7 eV.

These results lead us to two interesting afterthoughts. First, we have shown that the dispersion of the K plasmon linewidth is controlled by decay into electron-hole pairs involving unoccupied Kohn-Sham states lying in a narrow energy window —states which are commonly said to "not to mean anything." At a formal level, our finding is on a sound basis because of the TDDFT framework. At a physics level, we can rationalize this result as follows. We have found (by comparison with experiment) that the effects of the many-body kernel $f_{xc}$ are not important in the present problem; this is akin to stating that the excitonic electron-hole attraction does not play a role, in the present case.† In this sense, it is not too surprising that a one-particle picture may provide a good description of the degrees of freedom of the many-particle system probed in the plasmon decay process [these degrees of freedom (electron-hole states) involve no change in the number of particles, similarly to the case of optical absorption]. What *is* striking, and this takes us to the second comment, is that the Kohn-Sham one-particle "picture" is so successful in the present problem. Put differently, it is intriguing that the exchange-correlation "dressing" of the empty $d$-bands highlighted in Fig. 4a manifests itself directly in the overall shape of the dispersion curve for the linewidth of the K plasmon (Fig. 3b).‡

**DYNAMICAL STRUCTURE FACTOR OF ALUMINUM**

In the RPA, the dynamical density response for large wave vectors ($q \approx 2k_F$) corresponds to the excitation of particle-hole pairs which are nearly free (non-interacting). Furthermore, in the jellium model, Im $\mathbf{c}^{RPA} \to$ Im $\mathbf{c}^{(0)}$ for large $q$'s, where $\mathbf{c}^{(0)}$ is the Lindhard function —which in this limit is a *smooth function of frequency*. Now seminal IXSS experiments reported many years ago by Platzman and collaborators[21,22] showed that the response of relatively-simple systems such as Be, Al, Li, Si, and graphite, deviates significantly from this "expected" behavior. Indeed, the IXSS spectra for large wave vectors are dominated by an anomalous "two peak structure," whose discovery triggered a vast amount of work, which has persisted over the years, aiming at an understanding of the dynamical correlations in these systems —and, by extension, in solids in general.

With the advent of synchrotron sources, and the consequent improvement in energy and wave vector resolution (and photon fluxes), major advances have been reported over the last 10-15 years on the experimental front.[7-9,42,43] On the theoretical side,[44-50] until quite recently the thrust was directed almost exclusively towards the study of the effects of correlation for electrons in jellium.† It seems fair to state that this simple-model approach, which has featured treatments of the many-body problem of various degrees of sophistication, has not yielded spectra that one could consider to be faithful representations of the measured loss functions. This being the case, this approach remains inconclusive, in our view, since, in the absence of a small parameter for the interactions at play, comparison with the data plays a crucial role in the process of sorting out the relevance of proposed physical mechanisms. (For accounts of recent efforts within this approach, and differing views as to their relevance for an understanding of the data, see Refs. [7] and [9].)

---

† Now $f_{xc}$ does not have the direct diagrammatic interpretation of a vertex correction for the irreducible polarizability; thus our argument is only meant to convey a flavor for the physics at play.

‡ It is not obvious whether this effect of exchange and correlation is dealt with without uncontrollable double-counting in many-body models of interacting electrons, which typically include empirical band structures in the hopping term in the Hamiltonian.

It is only recently that *ab initio* calculations[11-20] have begun to change this state of affairs. The key reason behind this progress is the same "fortunate circumstance" we encountered above in the discussion of the collective mode in the alkali metals, where the Kohn-Sham single-particle response $c^{(s)}$ contained enough of the physics of the problem that we were close to the experimental data upon setting $f_{xc} = 0$ in Eq. (6). A similar situation is realized in the present case.

We recall that, in the first Born approximation, the double differential scattering cross section measured in IXSS is proportional to the dynamical structure factor $S(\vec{q};\omega)$, given by the equation

$$S(\vec{q};\omega) = -2\hbar V_N \, \mathrm{Im} \, c_{\vec{G},\vec{G}}(\vec{k};\omega), \qquad (7)$$

where $\vec{q} = \vec{k} + \vec{G}$. In Eq. (7) the wave vector *transfer* $\vec{q}$ is not restricted to the BZ. An appealing feature of IXSS is that it probes the electronic excitations over a wide range of wave vectors, from the small $q$'s for which the response is coherent (dominated by the plasmon) to the large $q$'s for which the response is incoherent (dominated by one-electron-like excitations), all the way to the Compton regime. For brevity, here we refer to a representative wave vector in the incoherent regime.‡ In Fig. 5a we discuss the $S(\vec{q};\omega)$ of Al for $|\vec{q}| = 1.7 k_F$; this case has been discussed at length by Platzman et al.,[7] Schülke et al.,[8] and Larson et al.[9,43] The IXSS data,[7] labeled $S_{\mathrm{exp}}$, are given by the empty circles; the most prominent feature of the measured loss function is the famous two-peak structure, whose minimum, centered at ~ 32 eV, is highlighted by a downward-directed arrow.

---

† The motivation for this widespread adoption of a simple-model approach was that the "double peak" (or one-peak, one-shoulder) loss structure was observed in systems with quite different band structures; furthermore, when the energy loss is measured in units of the Fermi energy, the double peak appears to suggest "universal" electron-gas behavior.

‡ It is noteworthy that the comparison which follows is made on the basis of theoretical and experimental spectra which are plotted in absolute units (by use of the *f*-sum rule); there is no lining up of peak heights, a procedure which would introduce a definite element of arbitrariness in the analysis. For example, if the comparison were based on lining up the height of the most prominent low-frequency peak, the overall agreement observed in Fig. 5a between theoretical and experimental spectra for frequencies above the two-peak structure would be destroyed. That agreement is an integral part of the reason why the conclusions we reach regarding the overall impact of the effects of the band structure are, in our view, quite compelling.[43]

Our analysis of the physics behind the data starts with a look at the loss function calculated using Im $c^{(s)}$ in Eq. (7) in place of Im $c$, i.e., the loss function $S_{KS}$ corresponding to the single-particle Kohn-Sham response (thick solid curve in gray tone.[15]) For reference, we also show the counterpart of $S_{KS}$ for electrons in jellium, $S_{Lindhard}$ (dots), which, when compared with $S_{exp}$, serves as a reminder of why the IXSS spectrum was originally described as being "anomalous."

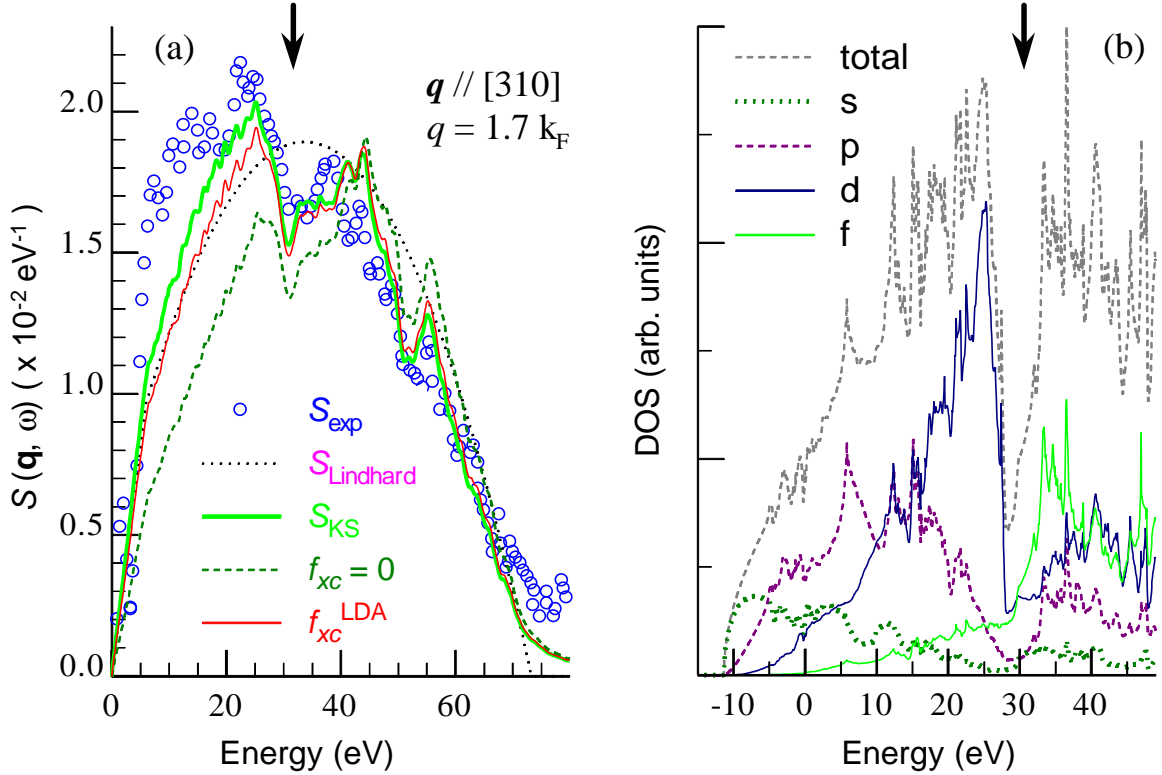

**Figure 5.** Explanation of the "double-peak" structure in the dynamical structure factor of Al. (a) The IXSS spectrum,[7] labeled $S_{exp}$ (empty circles), is compared with the loss function for Kohn-Sham electrons, labeled $S_{KS}$ (thick solid curve in gray tone), and with the loss function obtained from the solution of Eq. (4) for $f_{xc}=0$ (dashes) and for the LDA value of this kernel, $f_{xc}^{LDA}$ (thin solid line), respectively. For reference, the loss spectrum for non-interacting electrons in jellium with $r_s = 2.07$ (dots) is also shown. (b): Calculated DOS for Al; total DOS and contributions from states of *s*, *p*, *d*, and *f* symmetry, obtained in an LAPW band-structure calculation.[41] The zero of energy is the Fermi energy.

It is apparent that $S_{KS}$ reproduces the main features of $S_{exp}$ surprisingly well. In view of the exact nature of Eq. (4), this can only be true if, for the $\{\vec{q}, w\}$ subset of phase space explored in Fig. 5a, the many-body kernel $f_{xc}$ largely compensates for (i.e., cancels) the effect of the bare Coulomb interaction in the TDDFT density-response function $c$. Note that, in the light of this statement, the result for $S_{KS}$ in Fig. 5a can be viewed as providing a strong hint as to what one would really like to *predict* via an appropriate treatment of the dynamical correlations contained in $f_{xc}$.¶

---

¶ The exaggerated fine structure present in the calculated spectra will presumable be suppressed by dynamical-correlation effects left out in this calculation. Note also that $S_{KS}$ in Fig. 5a was obtained from a pseudopotential-based band structure;[15] thus it does not include the core excitations observed in the IXSS data at ~ 72eV.

A functional form for $f_{xc}$ which can be evaluated self-consistently with the electronic structure corresponds to the *static* LDA expression

$$f_{xc}(\vec{x},\vec{x}';\omega) = \frac{dV_{xc}[n](\vec{x})}{dn(\vec{x})}\delta(\vec{x}-\vec{x}'), \qquad (8)$$

where $V_{xc}[n](\vec{x})$ is the exchange-correlation potential entering the Kohn-Sham ground state problem within the LDA. That this approximate —and pragmatic— choice[51] for the many-body kernel leads to the type of cancellation effect suggested by the overall agreement between $S_{KS}$ and $S_{exp}$ is illustrated in Fig. 5a. First we note that, once we turn on the Coulomb interaction $v$ in Eq. (4), while still switching off the many-body kernel, the corresponding loss function, given by the dashes, and labeled $f_{xc}=0$ in Fig. 5a, worsens considerably, relative to experiment —the intensity on the low-frequency side is far too low. Nonetheless, the fine structure built into the Kohn-Sham response is still present in that loss function. Next, we have that, upon including the many-body kernel $f_{xc}$ into Eq. (4) according to the LDA prescription given by Eq. (8) —curve labeled $f_{xc}^{LDA}$ (thin solid line)— the quality of the low-frequency part of the calculated spectrum improves considerably. In fact, that result is quite close to the spectrum obtained for Kohn-Sham electrons, which is what we anticipated above.[†]

Of course, this near cancellation of the effects of the Coulomb interaction $v$ and the many-body vertex $f_{xc}$ in Eq. (4) does not hold "universally." For example, for wave vectors in the plasmon regime, Im $c^{(s)}$ bears little resemblance to the IXSS data[52] —thus the cancellation is a non-issue; numerically, the effect of $f_{xc}$ is minor, in this case.

The remaining question is the origin of the two-peak structure in $S_{KS}$.[¶] This point can be elucidated particularly convincingly with reference to the calculated DOS for Al, shown in Fig. 5b. It is, in fact, remarkable that, on the energy scale relevant to the present argument, the DOS is dominated by a deep indentation ("quasi-gap"), located at about the same energy as the minimum in the two-peak structure in $S_{KS}$ (note arrow in Fig. 5b). This wide quasi-gap in the DOS may seem to be a surprising feature of the band structure of Al, which is generally considered to be "jellium-like." However, there is no conflict. While, indeed, the bands near the Fermi surface are predominantly *sp*-like (see DOS in Fig. 5b), the excited-state band structure eventually differs significantly from simple parabolic bands. For example, there is the (2,0,0) zone-boundary gap, approximately 6 eV wide, centered at ~ 32 eV. Since the (310) direction which Fig. 5a refers to differs from the (100) direction by a relatively small angle, the final-state gap in the latter direction has a bearing on wave vector transfers along the former—thus the dip giving Im $c_{KS}$ its "two-peak" shape observed in Fig. 5a. But, why is the "gap" in the DOS (Fig. 5b) so large, if the crystal potential in Al is "weak?" The qualitative explanation for this apparently contradictory result is that, as shown in Fig. 5b, the angular-momentum content of the Bloch states is overwhelmingly *d*-like just below the gap at 32 eV, while above the gap both *d*-content and *f*-content dominate all other angular momentum components. Since the core states in Al contain neither $l=2$ nor $l=3$ components, the corresponding Bloch states near the gap are more exposed to the nuclear electrostatic potential than would be the case if these *l*-components contributed to the shielding between nucleus and valence states — thus the large gap at the (2,0,0) zone boundary, which is reflected in the pronounced dip in the DOS at 32 eV.‡ Naturally, this "final-state gap" depletes the available final states in the excitation process, and this gives rise to a dip in the loss function for Kohn-Sham electrons at about the same energy, as is clearly observed in Fig. 5a.

---

[†] Note that the effect of $f_{xc}$ observed in Fig. 5 is qualitatively similar to the "continuum exciton" effect discussed by Hanke and Sham in their study of the optical response of Si.[52]

[¶] See the more extensive discussion of this point presented recently by Larson et al.[43]

We summarize the above discussion by noting that the predominant two-peak structure in the IXSS spectrum of Al can be described pictorially as an indentation "carved out" from the Lindhard function (see Fig. 5a) because of the presence of a gap in the Kohn-Sham excited-state band structure. We emphasize again the sound basis which TDDFT gives to this argument. This framework has thus led us to the resolution of yet another "anomaly" in the experimental manifestation of the dynamical electronic response of a simple metal.

**EXCITATIONS IN REAL SOLIDS: CONSERVING DIAGRAMMATIC APPROACH**

The above calculations, while successful, cannot be considered to represent a complete theory; to reach that goal we must incorporate dynamical correlations according to Eq. (5) for the same electronic structure from which $c^{(s)}$ is obtained. In the context of the $S(\bar{q};w)$ of Al discussed above, an appropriate inclusion of the effects of a frequency-dependent $f_{xc}$ is expected to smooth out the exaggerated fine structure observed in Fig. 5a, while preserving the indentation due to the band gap at ~ 32 eV. On a wider scope, the dynamical effects of $f_{xc}$ must be accounted for if we are to study at an *ab initio* level systems which are characterized by complex electronic structures and non-trivial correlations near the Fermi surface ("strongly-correlated" electron systems).

In the case of atoms, progress has been reported recently in the implementation of the TDDFT linear-response scheme for an exchange-only functional.[24] In addition, significant advances have been made in the elaboration of theories of $f_{xc}$ which include correlation effects beyond adiabatic and local approximations;[55] however, these schemes have yet to be implemented for realistic representations of the band structure. Thus, for completeness, in the remainder of this article we sketch a diagrammatic alternative.

A rigorous formulation of the problem of one-particle-like excitations in a many-body system starts out from the Dyson equation, which we represent diagrammatically as

---

‡ It is interesting to note that this gap shows up clearly in the photoemission data of Levinson, Greuter and Plummer;[54] furthermore, their analysis of the angular momentum content of the states around the gap is in good agreement with the LAPW results of Fig. 5b.

$$\text{————} \quad = \quad \text{————} \quad + \quad \text{————}\!\!\!\bigcirc\!\!\!\Sigma^*\!\!\!\text{————} \quad , \tag{9}$$

where the thick line represents the actual one-particle Green's function $G(1,1')$, and the thin line represents the Green's function for "free" propagation in the LDA band structure, $G_{LDA}(1,1')$. All correlations beyond the LDA are formally built into the self-energy $\Sigma^*[G](1,1') = \Sigma[G](1,1') - (V_H(1) + V_{xc}(1))\delta(1-1')$, where $V_H(1)$ and $V_{xc}(1)$ are, respectively, the Hartree and exchange-correlation potentials entering the Kohn-Sham equation in LDA, and $\Sigma[G](1,1')$ is the self-energy for electron propagation in the "bare" external periodic crystal potential. Note that the cohesive effect of correlation (crystal binding) is built from the outset into the "unperturbed" Green's function $G_{LDA}(1,1')$.[†]

We proceed within the framework of the Baym-Kadanoff "conserving-approximation" scheme.[25] In the paper by Baym[26] it is shown that if an *approximate* model for the electron-electron interactions built into the self-energy is "$\Phi$-derivable," i.e., if there exists a functional $\Phi$, given by a set of Feynman diagrams composed of the bare interaction $v$ and the Green's function $G$, such that

$$\Sigma[G](1,1') = \frac{\delta\Phi}{\delta G(1',1)} , \qquad 1\!\!\bigcirc\!\!\!\Sigma\!\!\!\bigcirc\!\!1' \tag{10}$$

the ensuing Green's function is "conserving" —the self-consistent solution of Eqs. (9) and (10) fulfills all conservation laws (particle number, energy, momentum, angular momentum) *exactly*.[‡] In addition, "thermodynamic consistency" is ensured, which means, e.g., that different schemes for calculating the partition function give the same result.[¶]

A choice for $\Phi$ which incorporates the physics of screening due to the long-range Coulomb interactions is the shielded-interaction approximation (SIA),[25] which consists of the Hartree and Fock diagrams plus an infinite series of electron-hole bubbles made out of "exact" $G$'s —i.e., of the self-consistent solution of the Dyson equation for the chosen $\Phi$-functional— according to

$$\Phi_{\text{SIA}}: \quad \frac{1}{2}\ \text{[diagram]} - \frac{1}{2}\ \text{[diagram]} - \frac{1}{4}\ \text{[diagram]} - \frac{1}{6}\ \text{[diagram]} - \frac{1}{8}\ \text{[diagram]} - \cdots \tag{11}$$

---

[†] Comment on notation: The labels $1, 1'$ denote space-time points; the time variables are Matsubara times $0 \leq t, t' \leq \beta\hbar$. The arrow in all Green's functions indicates propagation from the space-time point on the right to the one the left; i.e., from $1'$ to $1$, in the case of $G(1,1')$. The arrow in the self-energy "cartoon" in Eq. (10) is simply a mnemonic device that helps decide the direction of the arrow for the propagators into which the former is to be inserted, as in Eq. (9). This convention about our cartoons (our nomenclature distinguishes cartoons from actual diagrams, such as the self-energy diagram in Eq. (12), etc.) is adhered to below in the case of other quantities which do not represent "propagation" of particles, such as the electron-hole interaction I, and the T-matrices $T^{ph}$ and $T^{pp}$.

[‡] We note that the $\Phi$-functional enters the exact expression for the thermodynamic potential $\Omega(T, \mu)$ obtained by Luttinger and Ward[56] from an analysis of the formal structure of many-body perturbation theory. In other words, Eq. (10) follows from the structure of the exact theory. The essential additional step taken by Baym was the demonstration that Eq. (10) ensures the fulfillment of the criteria for a conserving theory of $G$ put forth in the Baym-Kadanoff paper.[25]

[¶] This conclusion is non-trivial, since it holds for *approximate* (conserving) calculations, and errors propagate differently in different schemes for evaluating the same physical observable.[26,31]

From $\Phi_{\text{SIA}}$ we obtain, via Eq. (10), the following skeleton expansion for the self-energy:

$$\Sigma_{\text{SIA}}: \quad \text{[diagram]} - \text{[diagram]} - \text{[diagram]} - \text{[diagram]} - \text{[diagram]} - \cdots$$

$$= \text{[diagram]} - \text{[diagram]}, \tag{12}$$

which we have summed formally to all orders in $v$ via the introduction of the shielded interaction $W_{\text{SIA}}(11')$; this effective, or screened, interaction obeys the integral equation

$$W_{\text{SIA}}: \quad \text{[diagram]} = \text{[diagram]} + \text{[diagram]}. \tag{13}$$

We note that the label "SIA" attached to $W$ in Eq. (13) is meant to serve as a reminder of the fact that in the present case the shielded interaction is to be evaluated for the Green's function defined by this very set of equations.[†] If we were to change $\Phi$, $W$ would change. Of course, having started by choosing a subset of the exact $\Phi$-functional to generate $\Sigma_{\text{SIA}}[G_{\text{SIA}}]$, this approximation is automatically conserving.[‡]

We illustrate the physics contained in this scheme with reference to a recent calculation of the absolute band gap in Si.[27] In this context, it is relevant to note that the SIA is equivalent to Hedin's *GW* approximation,[28] which has proved extremely popular in *ab initio* calculations of quasiparticle band structures in metals and semiconductors, and more exotic forms of matter, such as the fullerines.[29,30,57] We work in the Bloch basis of the Kohn-Sham states, and turn Eqs. (9), (12), and (13) into matrix equations in the band

---

[†] One of the appealing features of the conserving method is the fact that self-consistency is built into it from the start —the self-energy arises naturally as a functional of the "exact" $G$. Of course, this adds to the computational complexity of the method, and helps explain why it has only been implemented recently (at the computational level),[27,31] starting with models of correlated electrons in two-dimensions, in the context of the high-$T_C$ superconductors.[58-61] The driving motivation for the resurgence of this method was the need to treat competing ordering mechanisms on an equal footing, self-consistency being crucial for even a qualitative description of the possibility of a change of phase.[62]

[‡] To be precise, the scheme assumes the presence of an external, fictitious, potential $U(11')$, which enters the equation of motion of $G(11')$; thus, $\Sigma$ depends on $U$ via its functional dependence on $G$. This technical point is important in the formulation of the method, as it ensures that the two-particle correlation function $L$, defined below, also fulfills the conservation laws.

index.[†] In Fig. 6 we show the spectral function $A_{jj}(\vec{k};\omega) \equiv -\pi^{-1} \operatorname{Im} G_{jj}(\vec{k};\omega)$ for the $\vec{k} j$ pairs defining the absolute band gap (we use the notation introduced in Eq. (6)).[‡] First we have the LDA; since this is a single-particle theory, the corresponding spectral function consists of delta functions, which, for ease of visualization, have been broadened numerically. The LDA band gap read off Fig. 6 (0.53 eV) differs from the experimental value (1.17 eV), by a factor of two. This was the original "band-gap problem."

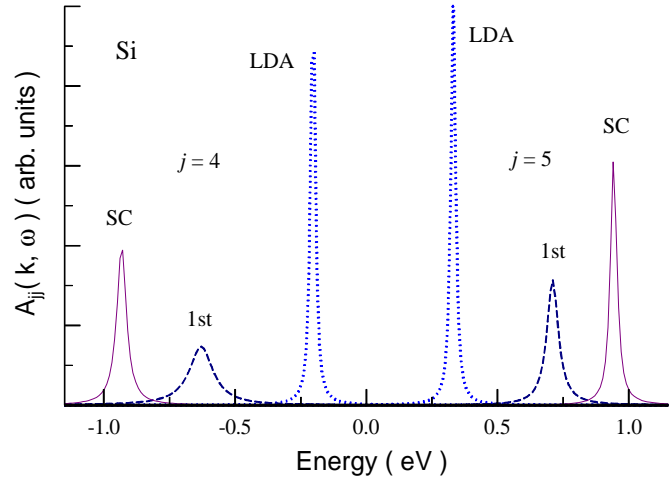

**Figure 6.** Impact of self-consistency on the absolute band gap in Si.[27] Shown is the spectral function $A_{jj}(k,\omega)$ for the states at the bottom ($j$=4) and the top ($j$=5) of the gap. Three cases are displayed, corresponding to the LDA, and to the first ("1st") and self-consistent ("SC") evaluations of the self energy. The zero of energy is the renormalized chemical potential for each case.

Next we have the gap corresponding to the quasiparticle band structure obtained from the self-energy $\Sigma_{SIA}[G_{LDA}]$, i.e., the solution of Eqs. (12) and (13) obtained with use of the LDA Green's function $G_{LDA}$. This calculation, which is not conserving, corresponds to the usual level of implementation of the *GW* approximation. The numerical value of the gap (1.34 eV) is now in good agreement with experiment (1.17eV). This is the usual success story — the band-gap problem appears to have been solved.

Finally, we perform the calculation at the conserving level; the Dyson equation is actually solved, and the new Green's function is used to recompute the self-energy iteratively until convergence. The gap obtained from $\Sigma_{SIA}[G_{SIA}]$ —the true *GW* band gap of Si— turns out to be too large (~ 1.9 eV), as can be surmised from Fig. 6. In fact, the experimental value of the gap is overestimated by about the same amount that the LDA underestimates it. Thus, *we have a band-gap problem once again*.[¶]

---

[†] Some details of the procedure we have implemented so far for the evaluation of all the ingredients of this system of equations are given elsewhere.[27] These include the evaluation of frequency convolutions which are insensitive to the cutoff which must necessarily be introduced, BZ sampling, number of bands kept in the evaluation of $G_{LDA}$, effects of finite temperature, analytic continuation to the real axis, etc.

[‡] The bottom of the gap corresponds to the fourth valence band, $j = 4$, at the $\Gamma$-point; the top of the gap (bottom of the conduction band) corresponds to the fifth band, $j = 5$, for $k \sim 0.8 \Gamma X$.

[¶] A similar situation is realized with regard to the calculated values of occupied bandwidths, which in the SIA are wider than the corresponding LDA values,[27,31] while experiment shows a narrowing, relative to the LDA.

It is clear that the treatment of correlation must be improved; the question is how to go about doing so. In the recent literature a number of suggestions related to this problem have been made.[30-38,63-68] It is important to bring these issues into the focus of the present discussion, since they have an impact on the development of a satisfactory description of quasiparticle band structure and density response.

For example, the issue has been brought up that in the SIA (that is, in the *GW* approximation, when implemented self-consistently) the electron-hole bubble entering Eq. (13) and the shielded interaction $W_{SIA}$ as well, "violate" the *f*-sum rule.[31,32] Here we stress the fact that both functions are simply building blocks for the evaluation of a conserving Green's function $G_{SIA}$, and of a conserving correlation function $L_{SIA}$ (see below); contact with experiment is made through *the latter* two functions. Thus, the fact that $W_{SIA}$ does not contain a plasmon pole, and, related to this, that it does not fulfill the *f*-sum rule,[‡] is not to be viewed as a "failure" of self-consistency; rather, in the context of the conserving scheme, this is a non-issue.[¶] We will return to this point below, when we introduce the irreducible polarizability for the conserving theory.

Related questions have been raised with regard to the satellite structure in $A_{jj}(\vec{k};w)$, whose spectral weight is strongly suppressed by the use of renormalized propagators in the SIA.[27,31] These issues, together with the fact that occupied bandwidths do not agree with experiment,[27,31] have prompted the suggestion that the effort invested in self-consistency may be "futile," since it leads to "inferior" results.[31] Similarly, it has been discussed whether "non-self-consistent *GW* is to be preferred over self-consistent *GW*.[63]"

With regard to the preceding question, we paraphrase it by asking whether it may be acceptable to ignore the fulfillment of basic conservation laws if this happens to lead to apparent agreement with experiment in some situations. Our viewpoint is that, if we are interested in understanding the electronic interactions at a fundamental level, what needs to be done is better many-body theory. And, in retrospect, why should one aspire to obtain, e.g., a "perfect" quasiparticle band gap while neglecting the contribution to the self-energy from the electron-hole attraction?[52,64-68]

It has also been suggested that the inclusion of vertex corrections should be pursued *instead* of adherence to self-consistency, when faced with the additional difficulty of carrying out these calculations for realistic models of the electronic structure.[31] This brings the problem into the realm of a rather large body of literature which focuses on the evaluation of the polarizability and dielectric function. Now, it is, of course, well-known that the effect of vertex corrections in the polarizability tends to compensate the effect of self-energy insertions in the same function. This "balancing" of both effects of correlation has been framed as a necessary condition for particle-number conservation.[33] In this light, the above SIA calculation (which includes a bubble which does not obey the *f*-sum rule) may seem to be flawed.[31] However, this is not the case: The solution $G(11')$ of the system of Eqs. (9), (12) and (13) is fully conserving, as we explained in detail above, even though the SIA self-energy functional does not include renormalized vertices. This point illustrates clearly why it is so important to agree on the language and framework that is being used.

---

[‡] We have checked numerically that this putative "violation" of the *f*-sum rule corresponds to a factor of 2 or more, depending on the wave vector; J. M. Sullivan, private communication.

[¶] Incidentally, note that the screening in the SIA is not the same as in the RPA —*W* is evaluated for different Green's functions in both cases. This distinction is essential in the conserving scheme.[25]

We turn next to the two-particle problem, which is the proper context for *f*-sum rule arguments. As noted below Eq. (13), the Baym-Kadanoff scheme requires the presence of an external, fictitious, non-local potential $U(11')$, which is introduced in order to "polarize" the system virtually. This artifice allows us to define the correlation function

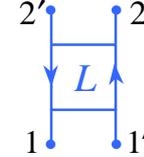

$$L(12;1'2') = \frac{dG(11')}{dU(2'2)}, \tag{14}$$

a subset of which,

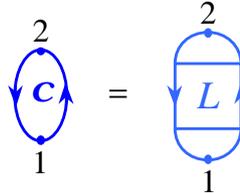

$$c(12) = L(12;1^+2^+), \tag{15}$$

yields (upon setting $U = 0$) the dynamical structure factor $S(\vec{q};w)$ according to Eq. (7). Now the direct evaluation of Eq. (14) is not very convenient, as it would require the explicit solution of the one-particle problem out of equilibrium. A better approach is to obtain $L$ from the Bethe-Salpeter equation it satisfies, namely,[25]

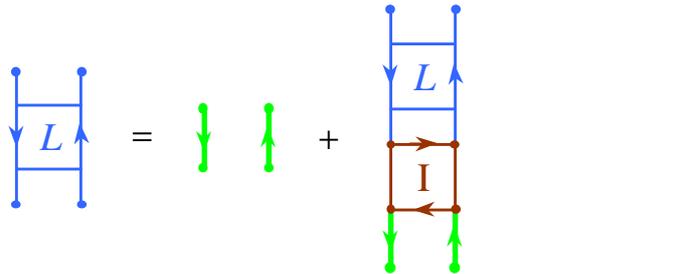

(16)

where now $U = 0$, and the effective "electron-hole" interaction $I(12;1'2')$ is given by

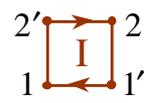

$$I(12;1'2') = \frac{d\Sigma(11')}{dG(2'2)}. \tag{17}$$

In principle, the physical content of the Bethe-Salpeter equation (16) is the same as that of the TDDFT response equation (4). If both schemes could be implemented exactly, they would yield the same loss spectrum $S(\vec{q};w)$. Of course, solving Eq. (16) is numerically much more demanding than solving Eq. (4), since the kernel $I(12;1'2')$ is a four-point function, while the kernel $f_{xc}(11')$ is a two-point function. The great conceptual merit of the conserving scheme is that, if the kernel $I(12;1'2')$ is generated according to Eq. (17), *on the basis of a conserving solution of the one-particle problem*, then $L(12;1'2')$ is also conserving —it fulfills all conservation laws in both $11'$ and $22'$ pairs of arguments; thus, in particular, it fulfills the *f*-sum rule. Note, again, that this exact fulfillment of the conservation laws for the two-particle response holds for an *approximate* functional $\Phi$.

It is useful to note that the Bethe-Salpeter equation can be iterated, and cast in the form of an integral equation for the two-point density-response function $c$ defined according to Eq. (15), namely

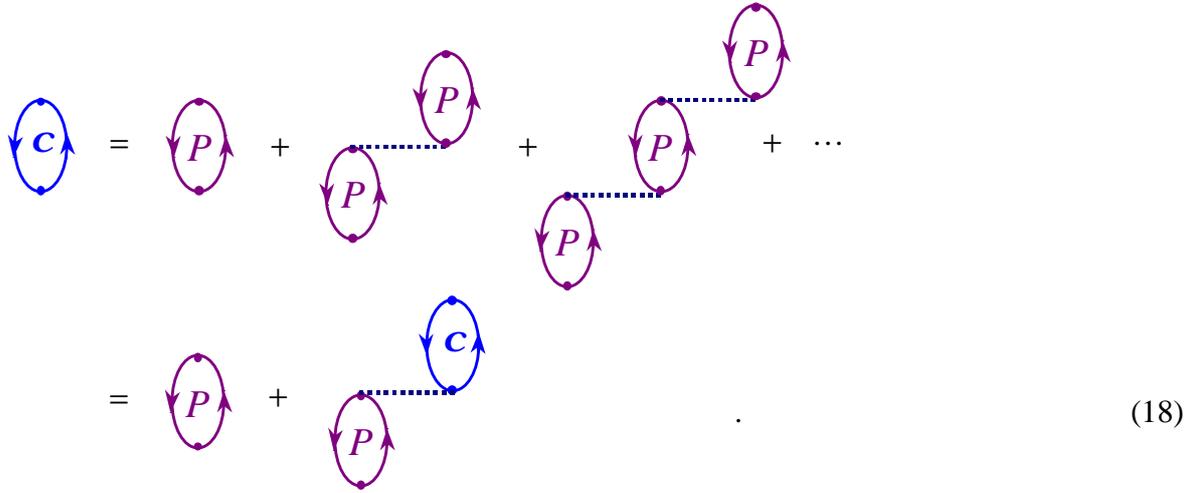

(18)

Equation (18), which is exact, is of the same *form* as Eq. (4).[†] Of course, all the complications of the many-particle processes contained in $I(12;1'2')$ are now hidden in the irreducible polarizability, $P(11')$. In fact, since both Eq. (18) and Eq. (4) yield the same response, we can readily write down an equation embodying a formal interrelation between $f_{xc}$ and $P$,

$$f_{xc}(11') = c_s^{-1}(11') - P^{-1}(11'),\qquad(19)$$

to which we will refer (briefly) below.

We illustrate the physics contained in the above scheme with reference to the SIA. Taking the functional derivative of $\Sigma_{\text{SIA}}$ required in Eq. (17) yields the following result for $I_{\text{SIA}}$:

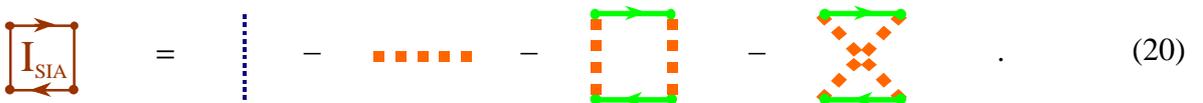

(20)

---

[†] Equation (18) can be derived by the usual many-body perturbation theoretic techniques.[68] The present approach has the added advantage that it yields a response $c$ which is guaranteed to be conserving.

The *structure* of the first two contributions to this effective electron-hole interaction corresponds to the so-called time-dependent screened Hartree-Fock approximation; that approximation is on a different footing —it is *not* conserving. The last two terms in Eq. (20) have traditionally been neglected in *ab initio* work for realistic models of solids; however, their inclusion is necessary if we seek a conserving response, $c_{SIA}$. Substituting Eq. (20) into Eq. (16), we obtain the counterpart of Eq. (18) for the specific case of the SIA, whose solution is given, in symbolic notation, by $c_{SIA} = [1 - v P_{SIA}]^{-1} P_{SIA}$, in terms of the irreducible polarizability $P_{SIA}$, defined by the equation

$$P_{SIA} = \bigcirc + \bigcirc\!\!\bigcirc + \bigcirc\!\!\bigcirc\!\!\bigcirc + \bigcirc\!\!\bigcirc\!\!\bigcirc + \cdots + \bigcirc\!\!\bigcirc\!\!\bigcirc + \cdots \text{ (mixed terms)}$$

$$+ \bigcirc\!\!\bigcirc + \bigcirc\!\!\bigcirc\!\!\bigcirc + \bigcirc\!\!\bigcirc\!\!\bigcirc + \cdots$$

$$+ \cdots \tag{21}$$

We emphasize that the first term on the right-hand-side of Eq. (21) is precisely the dressed electron-hole bubble which enters Eq. (13) for the self-energy $\Sigma_{SIA}$. It should now be apparent why we stated above that it was a non-issue for this bubble to "fail" to fulfill physical constraints, such as the *f*-sum rule.

In order to visualize the complex interactions built into $P_{SIA}$, we have grouped Eq. (21) in columns, each column beyond the first one having the structure of a ladder whose rungs are given by the respective terms in Eq. (20). The first such shielded ladder, which is traced to the second term in Eq. (20), has the structure of the ladders originally approximated by Hubbard via the introduction of a "local-field factor" (although in that case the ladders were not dynamically screened). The third and fourth columns in Eq. (21) correspond, respectively, to the last two terms in Eq. (20); these ladders are "fluctuation diagrams."[70] The remaining columns correspond to more complicated fluctuations containing a mixture of the processes which originate in the last three terms in $I_{SIA}$.[†]

---

[†] Note that, in order for $c_{SIA}$ to be strictly conserving, *all* the processes contained in Eq. (21) must be incorporated. Thus, in spite of the apparent simplification introduced on going from Eq. (16) to Eq. (18), we advocate the (very demanding) direct numerical solution of the Bethe-Salpeter equation.

We note that the first four diagrams in the first row of Eq. (21) correspond to the polarization diagrams whose contribution to $f_{xc}$, to the lowest order in $W_{RPA}$, was elucidated by Langreth and Vosko.[70][†] In the light of Eq. (19), that work provides the beginnings of an explicit interrelation between the two approaches to correlated-electron response we have outlined here. We hope to explore this issue further in future work.

In addition, we note that the diagrams highlighted in the previous paragraph are responsible for the leading correction to the RPA density-response function for the high-density electron gas.[71] Furthermore, each of the these diagrams has non-integrable Fermi-surface-related singularities which become integrable upon summing them all up;[70] the weight of the integrable singularity increases with $r_s$.[‡] Since the last two diagrams in this set originate from the last two terms in Eq. (18), it seems justified to assert that the common practice of neglecting these electron-hole interactions is unwarranted —of course, their inclusion is also required from the point of view of the conservation laws.

Finally, we stress the fact that the structure of Eq. (21) is consistent with the rules which have been discussed by Mahan for balancing vertex corrections and self-energy insertions in the irreducible polarizability, in compliance with a Ward identity.[33] However, it is extremely important to note that in the conserving method the polarization function is "predetermined" by the choice of the $\Phi$-functional, and by the definition of the electron-hole interaction according to Eq. (17), *which is to be implemented following the conserving solution of the one-particle problem* —as we have illustrated with our analysis of the polarization $P_{SIA}$, which follows from our choice of $\Phi$-functional, $\Phi_{SIA}$ (Eq. (11)). No such balancing of vertex corrections and self-energy insertions is imposed on the solution of the one-particle problem in order for $G(11')$ to be conserving —which is why this "balance" has no bearing on the conserving nature of $G_{SIA}$, or on the result for the Si band gap presented in Fig. 6, as we argued above.

We proceed to summarize the above presentation. First, we have discussed the implementation of the conserving scheme at the level of the one-particle problem. For the specific choice of $\Phi$-functional given by Eq. (11) —the SIA— we have noted that the self-consistent solution of Eqs. (9) and (12) yields a value of the absolute band gap in Si which disagrees with experiment. *There is a band gap problem, once again.*

One possible scenario is that there is some conceptual flaw with the SIA, which, as we have noted repeatedly, is the same as the *GW* approximation —when implemented self-consistently, which was the original intent in the Hedin scheme (although his method was not concerned with conserving approximations per se). We have considered this scenario at some length, and we hope that our discussion of the conserving scheme, including the crucial interrelation between the one-particle and two-particle problems, has made a clear case for the fact that the SIA self-energy functional is free of a variety of objections which have been raised in the literature —in the related context of the effect of self-consistency— with regard to sum rules, balance of vertex corrections and self-energy insertions, etc.

---

[†] $W_{RPA}$ has the structure of Eq. (13), but the bubble is evaluated for Hartree Green's functions.

[‡] However, note that in the analysis of Langreth and Vosko the diagrams are evaluated with bare propagators. Nonetheless, their results are very suggestive of the importance of these diagrams for finite values of $r_s$.

Another scenario —the one we are advocating herein— is that a better many-body calculation of the gap is needed. The apparent agreement with experiment obtained at the non-conserving level results from a cancellation of errors: lack of self-consistency (violation of the conservation laws), and absence of short-range correlation effects. In retrospect, this "new band gap problem" is not too surprising. In effect, it has been shown that the excitonic electron-hole attraction plays a significant role in the optical absorption process in Si.[52,64-66] In view of the results of Fig. 6, we expect that the electron-hole interaction must contribute to the value of the *quasiparticle* band gap.

The effects of the particle-hole interaction may be incorporated into the one-particle problem in terms of a $\Phi$-functional (displayed explicitly in the Appendix, where it is labeled $\Phi_T^{ph}$) which, via Eq. (10), generates a self-energy $\Sigma_T^{ph}$ of the form

$$\Sigma_T^{ph} = - \;\; T^{ph} \quad , \tag{22}$$

where the particle-hole T-matrix, $T^{ph}$, is the solution of the integral equation

$$T^{ph} = \quad - \quad T^{ph} \quad . \tag{23}$$

We note that, when adding the self-energy processes contained in Eq. (22) to those contained in Eq. (12), care must be exercised in order to avoid double counting the second-order diagram.

Another type of short-range correlation process left out in the SIA involves short-range particle-particle interactions (including exchange processes), which are known to be essential in the low-density limit.[72] These processes can be incorporated into a $\Phi$-functional (again, this functional is displayed explicitly in the Appendix, where it is labeled $\Phi_T^{pp}$) which, via Eq. (10), generates a self-energy $\Sigma_T^{pp}$ of the form

$$\Sigma_T^{pp} = - \;\; T^{pp} \quad , \tag{24}$$

where the particle-particle T-matrix, $T^{pp}$, is the solution of the integral equation

$$\boxed{T^{pp}} \;=\; \boxed{\phantom{xx}} \;-\; \boxed{\times} \;-\; \boxed{\phantom{x}\,T^{pp}} \quad . \tag{25}$$

Again, double counting of the second-order term must be avoided, when adding this new contribution to the self-energy. Equations (12) and (25) provide a good interpolation between short-range and long-range correlations (or between low-density and high-density physics) in an electron gas.[46,48,73] Furthermore, the inclusion of the correlations contained in Eqs. (22) and (25) is expected to play a role in the description of more strongly-correlated materials, such as the transition metals, and their oxides.[67]

**APPENDIX**

An important set of particle-hole interactions, absent in the SIA, is represented by the $\Phi$-functional

$$\Phi_T^{ph}: \quad -\frac{1}{4}\;\bigcirc\;+\;\frac{1}{6}\;\bigcirc\;-\;\frac{1}{8}\;\bigcirc\;+\;\cdots$$

which, upon taking the functional derivative required in Eq. (10), generates a series of "particle-hole" self-energy ladders,

$$\Sigma_T^{ph}: \quad -\;\bigcirc\;+\;\bigcirc\;-\;\bigcirc\;+\;\cdots$$

$$= \;-\;\boxed{T^{ph}}$$

which we have formally summed to all orders via the definition of the particle-hole T-matrix, $T^{ph}$, which is to be obtained by solving Eq. (23).

A family of particle-particle interactions (including exchange processes), also absent in the SIA, is represented by the $\Phi$-functional

$$\Phi_T^{pp}: \quad -\frac{1}{4}\;\bigcirc\;+\;\frac{1}{6}\;\bigcirc\;-\;\frac{1}{8}\;\bigcirc\;+\;\cdots$$

$$\quad +\;\frac{1}{4}\;\bigcirc\;-\;\frac{1}{6}\;\bigcirc\;+\;\frac{1}{8}\;\bigcirc\;-\;\cdots$$

which, upon taking the functional derivative required in Eq. (10), generates a series of "particle-particle" self-energy ladders,

$\Sigma_T^{pp}$ : [diagrammatic expansion] $= -T^{pp}$

which we have formally summed to all orders via the definition of the particle-particle T-matrix, $T^{pp}$, which is to be obtained by solving Eq. (25).

## ACKNOWLEDGEMENTS

This work was supported by National Science Foundation Grant No. DMR-9634502 and the National Energy Research Supercomputer Center. Oak Ridge National Laboratory is managed by Lockheed Martin Energy Research Corp. for the Division of Materials Sciences, U.S. DOE under contract DE-AC05-96OR2464.